\documentclass[a4paper,11pt]{article}
\usepackage{bm,mathrsfs,amsmath,amsthm,amssymb,eepic,amscd,longtable,array,graphicx}
\usepackage[dvips]{color}
\newcommand{\nc}{\newcommand}
\nc{\rnc}{\renewcommand}
\nc{\nn}{\nonumber}
\nc{\del}{{\partial}}
\rnc{\Im}{{\rm{Im}\,}}
\rnc{\Re}{{\rm{Re}\,}}
\nc{\db}{\displaybreak[0]\\}
\nc{\bra}{\langle}
\nc{\ket}{\rangle}

\nc{\lam}{\lambda}
\nc{\g}{{\mathfrak{g}}}
\nc{\zb}{\bar{z}}
\nc{\hb}{\bar{h}}
\nc{\J}{\mathcal{J}}
\nc{\su}{\widehat{\mathfrak{su}}(2)_k}

\nc{\tcr}{\textcolor{red}}

\numberwithin{equation}{section}
\numberwithin{lemma}{section}
\numberwithin{proposition}{section}
\numberwithin{theorem}{section}
\numberwithin{corollary}{section}
\numberwithin{conjecture}{section}

\begin{document}%
%
\title{Time reversing procedure of SLE and 2d gravity}

\author{
\\
Yoshiki Fukusumi$^1$\thanks{E-mail: y.fukusumi@issp.u-tokyo.ac.jp} 
\\\\
%
{\it Institute for Solid State Physics, University of Tokyo}\\
{\it Kashiwa, Chiba 277-8581, Japan} \\
}
\maketitle
%

%
\begin{abstract}
We analyse the time reversing procedure of  Schramm Loewner evolutions and
its relation to Liouville field theory or 2d pure gravity. It enables one to obtain martingale
observables by the calculation of Liouville field theory correlation functions.
  
\end{abstract}
%


%
\section{Introduction}
There are a wide variety of connections between Schramm Loewner
evolutions and conformal field theories. One of the most direct and physical
methods to construct SLE and CFT correspondence is the construction of martingale observables
whose expectation values coincide with correlation functions of CFT \cite{FW,BB,BBK}. This martingale
condition ensures CFT calculations of SLE probabilities such like left passive
probability or connectivity weight. 
Begining from the work by D. Bernard and M. Bauer, W. Wernar and R. Friedrich, there are a lot of
variant of that connection. Recently generalized SLEs are constructed by field theoretic
argument and a lot of them are consistent with Monte Carlo simulation
of 2d lattice models \cite{FSKZ}.  
 
Moreover by the series of works by Duplantier and Sheffield, it was widely recognized that
SLE martingale is closely related to the random measure of 2d gravity.\cite{DS}
They constructed random measure of 2d gravity
by SLE on free boundary condition. It is different from usual SLE/CFT correspondence in that SLE describes interface
or boundary condition changing operator. And series of their papers, they showed conformal
invariance of such one point functions.

In this paper, we show multipoint functions of Liouville field theory are constructed as the martingale
distribution functions of the time reversed SLE. Our formulation is based on previous work of Duplantier and
Sheffield, but we assume reverse flow of SLE should be treated as chordal version of whole plane SLE.
If our formulation are valid, we can get boundary Liouville field theory as martingale distribution
functions of generalized  SLE.

\section{Time reversed SLE}
What we call time reversed SLE is the same as the reverse (or backward ) SLE considered
in 2d gravity coupled with matter. To stress the role of time, we call it as time reversed SLE in this paper.
It is defined by formal time reversal of SLE and colsely related to
the reversability of  SLE. First we introduce the chordal SLE $g_t :H\setminus K_t \rightarrow H$ as,
\begin{equation}
dg_{t}\left(z\right) =\frac{2dt}{g_{t}\left( z \right)-\xi _t}, \xi _t=\sqrt{\kappa} B_{t},
\end{equation}
and it satisfies the initial conditions $g_0 (z)=z$.
For the later discussion, the property of the tip of SLE curve is
important i.e. $g_t (\gamma _t)=\xi _t$.   

For introducing reverse time SLE, we think about the SLE process
from $t=0$ to $t=T$.  
Then one can obtain the data of ${( B_t) }_{t=0}^{t=T}$.
In this situation, one can define time reversed Loewner map,
\begin{equation}
dg_{T-t} (z)=\frac{-2dt}{g_{T-t} (z)-\xi _{T-t} }.
\end{equation}
For convenience we omitted the time $T$ and transform $t\rightarrow -t$.This process
is defined by the condition $(B_t)^T_t$. This $B_t$ is not
stochastic object in this stage but the consequence of Brownian motion
and it is fractal object.
So we can assume,
\begin{equation}
(dB_t)^2 =dt, a.e.
\end{equation}
Therefore it is natural to introduce time reversed or backward SLE,
\begin{equation}
dg_{t} (z)=\frac{-2dt}{g_{t} (z)-\xi _{t} }.
\end{equation}
with the same initial condition as that of original SLE.

In the previous paper by Duplantier, the radial stochastic version of this theory 
is confirmed to become a uniformizing map \cite{DHLZ}.
However the chordal version of this SLE is a map $H\rightarrow H\setminus K_t$.
In the section 4, we show this map has the Liouville theory correlation function as
the martingale observable.

\section{Very short review of the generalized Liouville Field theory}
In this section, we review the genralized Liouville field theory which
includes minimal matter CFT and Liouville field theory. In general
one can define these CFT by Hilbert space and OPE or commutation relation
of operators.
(For complete discussion, please see the paper by Ribault \cite{Rib}and its reference)
In SLE/CFT sence, these structure of CFT naturally generates martingale observables.
 
The central charge and conformal dimension of the theory is
given by,
\begin{equation}
c=1+6Q^2,\
Q=b+1/b.
\end{equation} 
This parameter $b$ is a complex number which determines the theory.
The theory becomes matter CFT if $b$ is pure imaginary, and becomes
pure gravity or Liouville field Theory if $b$ is real.

One of the most characteristic points of Ribault's generalized Liouville field theory is
the assumption of the existence of degenerate fields. 
For the calculation of correlation functions, degenerate fields are useful and
fundamental to define CFT.
In bosonisation or Wakimoto free field representation scheme,
each contour of screening field corresponds to the solution of the 
differential equation derived from degenerate conditions.

The conformal dimensions of degenerate Fields are given by
\begin{equation}
h_{(r,s)}=\alpha_{(r,s)}(Q-\alpha_{(r,s)}),\
\alpha_{(r,s)}=\frac{Q}{2}-\frac{1}{2}\left( br+b^{-1} s \right)
\end{equation}

For the later convenience we introduce the level 2 null vectors of
generalized minimal models.
\begin{equation}
\left( b^2L_{-1}^2+L_{-2}\right) \phi _{(1,2)}=\left( \frac{1}{b^2}L_{-1}^2+L_{-2}\right) \phi _{(2,1)}=0.
\end{equation} 
What is important for the later discussion is that the sign of $b^2$ can be changed
by the transformation $b'=ib$ and that transformation changes central charge $c'=26-c$.
This dual transformation relates pure gravity and matter CFT, which can couple. 

If we assume mode expansion of energy momentum tensor $T(z)=\sum_{n}{L_n z^{n+2}}$,
and operator expansion of fields, the degenerate condition transforms to the following differential equation,
\begin{equation}
\left( b^2\partial^{2}_{z}+\sum_{\alpha}{\left( \frac{h_{\alpha}}{(z_\alpha-z)^2}-\frac{\partial_{z_\alpha}}{z_\alpha -z}\right)}\right)\bra \Pi_{\alpha}{X({z_\alpha})\phi_{(1,2)}(z)} \ket=0,
\end{equation} 
with $\Pi_{\alpha}{X({z_\alpha})}=\Pi_{\alpha}\phi_{h_\alpha}(z_\alpha)$ and $\phi_{h_\alpha}(z_\alpha)$ is primary
field with conformal dimension $h_\alpha$.
\section{Martingale observable of time reversed SLE}
In this section, we show the correlation functions of Liouville field theory
are consistent with the time reversed SLE.

First we define observables. We define the
primary fields with conformal dimension $h$,
\begin{equation}
\phi_h(z)=(g'(z))^h\phi_h (g_t(z)).
\end{equation}
Moreover we assume interfaces are described by boundary condition changing
operator $\psi$. In this stage, we don't assume its conformal dimension of this operator.
What should be done is detecting the conformal dimension of this
b.c.c. operator on each case.
We consider following observable on the upper half plane $H$ in analogy with usual SLE,
\begin{equation}
F(y_{\alpha},\xi, t)=\prod_{\alpha}(g'(y_{\alpha}))^{h_{\alpha}}\frac{\bra \prod_{\alpha} \phi_{h_\alpha}(g_t(y_{\alpha}) )\psi(\infty)\psi(\xi_t)\ket_H}{\bra\psi(\infty)\psi(\xi_t)\ket_H}.
\end{equation}
For this curve goes to $\infty$, we took boundary condition changing operator at $\infty$.
Unfortunately, $g_t$ maps $H$ to $H\setminus K_t $, usual SLE/CFT correspondence
is not valid for chordal SLE case. However, in radial SLE, such difficulty does not appear. 

Then one can derive derivative of primary fields under time reversed SLE,
\begin{equation}
d\phi_{h _\alpha}(y_\alpha)=
-2dt {(g_t '(y_\alpha))}^{h_{\alpha}}( \frac{\partial_{y_\alpha}}{g_t (y_{\alpha})-\xi_t}-\frac{h_{\alpha}}{(g_t (y_{\alpha} 
)-\xi_t)^2})\phi_{h_{\alpha}}(g_t (y_{\alpha})).
\end{equation}
By Ito calculus, derivative of boundary field is,
\begin{equation}
d\psi (\xi_t)=d\xi_t \psi '(\xi_t)+\frac{\kappa}{2}dt\psi ''(\xi_t).
\end{equation}

Therefore the total derivative of the observable is,
\begin{equation}
\begin{split}
\prod_{\alpha}{g_t(y_\alpha)^{-h_\alpha}} dF(y_{\alpha},\xi, t)  
= &\left(dt\left(\frac{\kappa}{2}\partial^{2}_{\xi_t} -2\sum_{\alpha}{ \left(\frac{\partial_{y_\alpha}}{
g_t (y_{\alpha})-\xi_t}-\frac{h_{\alpha}}{(g_t (y_{\alpha} )-\xi_t)^2}\right)}\right)
+d\xi_t \partial_{\xi_t})\right) \\
& \bra \prod_{\alpha}{\phi_{h_\alpha}(g_t (y_\alpha))}\psi(\xi_t) \psi(-\infty) \ket .
\end{split}
\end{equation}
Therefore the martingale condition is nothing but null vector equation of Liouville gravity in the
previous section with $b^2=\kappa/4$, $\psi=\phi_{(1,2)}$. If there is no other freedom of Brownian motion like ones on Lie group manifold \cite{BGLW},
we can obtain the time reversed SLE observables by calculating Liouville field theory correlation function.  
Moreover this Liouville field theory has central charge  $c_L =1+6\left( \sqrt{\frac{\kappa}{2}}+ \sqrt{\frac{2}{\kappa} } \right)^{2}$ and that is interesting original SLE is related to matter CFT with central charge $c_M=1-6\left( \sqrt{\frac{\kappa}{2} } -\sqrt{\frac{2}{\kappa} }\right)^{2}$. These central chages satisfies the relation  $c_L +c_M =26$.
This relation is very famous as the coupling condition of minimal matter CFT and 2d gravity.

\section{An explicit form of maritingale observable}
As we have shown in the previous sections we can get the martingale observables
in the form of Liouville Field theory correlation functions.
In this section, we give the most simplest observables,
\begin{equation}
\bra \phi_{(1,3)} (z)\ket_{H}=(g'_t (z))^{-1-\frac{8}{\kappa^2}}(g_t (z)-\xi_t)^{-1-\frac{8}{\kappa^2}}
\end{equation}
It is the almost same form of observable which is in the paper by Duplantier and Sheffield.

What Duplantier and Sheffield was proved was following,
\begin{equation}
(D,\phi)_{H}=(\psi(D),\phi)_{H\setminus K_t}.
\end{equation}
Their right hand side should be calculated by free boundary condition
on $H$. 

On the other hand our formulation is consistent with BCFT picture.
\section{Some conjectures}

At least radial SLE case, there exists whole plane SLE which corresponds to our reverse SLE
interpretation, in which the curve goes to outside of domain\cite{DHLZ} and map the tip to boundary
Brownian motion.
Hence we expect we can get martingale observable of whole plane SLE as the correlation functions of
Liouville field theory.

Based on this observation,
we propose a process which may be related to 2d gravity coupled with matter. 

The stochastic process is described by following map $h_t\circ g_t:H\setminus K_{t,2}\rightarrow H \setminus K_{t,1}$.
\begin{equation}
dh_t(z)=\frac{-2dt}{h_t(z)-\xi_{t,1}},
\end{equation} 
\begin{equation}
dg_t(z)=\frac{2dt}{g_t(z)-\xi_{t,2}},
\end{equation}
with $h_t:H\rightarrow H\setminus K_{t,1}$, $g_t:H\setminus K_{t,2}\rightarrow H$,
and initial condition $g_0=h_0=id$ and the tips are given by 
 $g_t(\gamma_{t,1})=\xi_{t,1} $.
$\xi_{t1}$ and $\xi_{t,2}$ are Brownian motions.

Moreover, one can define multiple time reversed SLE by the almost same procedure for usual
multiple SLE. In the proceeding paper, we will treat that generalization.

Finaly, if we assume this time reversing procedure changes the matter to the corresponding
gravity, we can get corresponding gravity for usual Wess-Zumino-Witten models.
For example, it may be interesting to check coupling condition for $SU(2)_k$ WZW model\cite{ABI}.
It is known this model can couple to $SL(2,R)$ WZW model with $c_{SU(2)}+c_{SL(2)}=6$\cite{MM}.
Moreover, although SLE for $Z_n$ parafermion is constructed by R. Santachiara and M. Picco \cite{San,PS,PSS},
the gravity which can couple to $Z_n$ parafermion is not known. Therefore it may be interesting
to check the null vector condition for $SL(2,R)_k/U(1)_k$ gravity.
\section{Conclusion}
In this paper, it was shown that the martingale observable of time reversed SLE 
can be constructed from the correlation functions of Liouville field theory or 2d pure
gravity.
That means one can obtain pure 2d gravity CFT by the calculation of the distribution
function of SLE. 
As a consequence one can obtain Liouville field theory with $c_L =1+6\left( \sqrt{\frac{\kappa}{2}}+ \sqrt{\frac{2}{\kappa} } \right)^{2}$ and minimal matter CFT with $c_M=1-6\left( \sqrt{\frac{\kappa}{2} } -\sqrt{\frac{2}{\kappa} }\right)^{2} $. If one fix $\kappa$, there is the relation $c_L +c_M =26$. That means formal time reversing procedure relates matter and gravity which can couple \cite{RZ}.

We hope our formulation fulfills the gap between SLE on minimal matter CFT and
reverese SLE on 2d gravity coupled to minimal matter.
In the forth coming paper we will try to construct multiple reverse SLE on Liouville
field theory\cite{BBK}.

Just before the end of this work, I noticed the work which is confirming
my observation for radial case \cite{Alek}.
    
\section{Acknowledgement}   
First I would like to thank Sylvain Ribault for the beautiful lecture
on "Quantum integrable systems conformal field theories and
stochastic processes" at Cargese.
I also thank Kazumitsu Sakai for helpful comment
and Yuji Tachikawa, Masaki Oshikawa, Guillaume
Remy for discussion
and comments. 

\appendix
\def\thesection{\Alph{section}}
\def\reference{\relax\refpar}

\section{Radial case}

In this section, we will show whole plane SLE $g_t$ satisfies martingale for Liouville
field theory. The monotonicity of the map corresponding to $g_t^{-1}$ has already
shown. Therefore we assume $g_t$ as a map to unit disc.  
The same procedure was already done by O. Alekseev in the context of Laplacian growth and Liouville gravity.

First we transform the whole plane SLE as a map to upper half plane.
The same procedure was done in radial SLE.
The stochastic equation becomes,
\begin{equation}
dg_t(z)=-\frac{1+g_t (z)^2}{2}\frac{1+\eta_t g_t (z)}{g_t (z)-\eta_t}dt, \ \eta_t =\tan \xi_t
\end{equation}

We heavily use the group theroretic discussion of M. Bauer and D. Bernard \cite{BBrad}.
Then the group theoretic representation of this whole plane SLE on the upper half
plane is,

\begin{align}
H_t^{-1}dH_t=dt\left( 2W_{-2}+\frac{\kappa}{2}W_{-1}^2\right)+d\xi_{t}W_{-1}, \\
W_{-1}=\frac{1}{2}\left( L_{-1}+L_{1}\right), W_{-2}=\frac{1}{4}\left( L_{0}+L_{-2}\right).
\end{align}
Then by the stochastic Ito calculus, we can obtain the martinagale observable,
\begin{equation}
e^{-h_{r}t}H_t |\phi_{(1,2)}\ket, h_{r}=-\frac{(2+\kappa)(6+\kappa)}{8\kappa}.
\end{equation}
This condition can be calculated by
\begin{equation}
\left(2W_{-2}+\frac{\kappa}{2}W_{-1}^{2}\right)|\phi_{(1,2)}\ket=-\frac{(2+\kappa)(6+\kappa)}{8\kappa}|\phi_{(1,2)}\ket .
\end{equation}


\end{document}